\documentclass[prb,preprint,showpacs,preprintnumbers,amsmath,amssymb,superscriptaddress]{revtex4-2}

\usepackage{graphicx}
\usepackage{dcolumn}
\usepackage{bm}
\usepackage{amsthm}
\usepackage{amsmath}
\usepackage{amssymb}
\usepackage{xcolor}

\begin{document}

\title{On the angular anisotropy of the randomly-averaged magnetic neutron scattering cross section of nanoparticles}
 
\author{Michael P.\ Adams}\email[Electronic address: ]{michael.adams@uni.lu}
\affiliation{Department of Physics and Materials Science, University of Luxembourg, 162A~avenue de la Faiencerie, L-1511~Luxembourg, Grand Duchy of Luxembourg}

\author{Evelyn Pratami-Sinaga}\email[Electronic address: ]{evelyn.sinaga@uni.lu}
\affiliation{Department of Physics and Materials Science, University of Luxembourg, 162A~avenue de la Faiencerie, L-1511~Luxembourg, Grand Duchy of Luxembourg}

\author{Andreas Michels}\email[Electronic address: ]{andreas.michels@uni.lu}
\affiliation{Department of Physics and Materials Science, University of Luxembourg, 162A~avenue de la Faiencerie, L-1511~Luxembourg, Grand Duchy of Luxembourg}


\begin{abstract}
We calculate the magnetic SANS cross section of dilute ensembles of uniformly magnetized and randomly-oriented Stoner-Wohlfarth particles using the Landau-Lifshitz equation. The focus of our study is on the angular anisotropy of the magnetic SANS signal as it can be seen on a two-dimensional position-sensitive detector. Depending on the symmetry of the magnetic anisotropy of the particles (\textit{e.g.}\ uniaxial, cubic), an anisotropic magnetic SANS pattern may result, even in the demagnetized state or at the coercive field. The case of inhomogeneously magnetized particles and the effects of a particle-size distribution and interparticle correlations are discussed. 
\end{abstract}

\date{\today}

\maketitle

 
\section{Introduction}

In a magnetic SANS experiment, the angular intensity distribution of the scattered neutrons on the two-dimensional position-sensitive detector usually provides the first information on the magnetic microstructure of the sample under study. When an external magnetic field $\mathbf{B}$ is applied and varied during the experiment, such images can yield useful information on the degree of magnetic saturation (at large fields), on the presence of clover-leaf-shaped angular anisotropies (at intermediate fields), or whether or not the magnetic moments are randomly distributed (at remanence or at the coercive field). The angular anisotropy of the magnetic SANS cross section can have many origins~\cite{michelsbook}, \textit{e.g.}\ (i)~it can be due to the dipolar nature of the interaction between the magnetic moment of the neutron and the magnetic moment that is formed by the unpaired electrons of the sample, (ii)~it might be related to the magnetic interactions within the sample such as the magnetodipolar energy between magnetic moments, anisotropic exchange interaction, or magnetic anisotropy, or (iii)~it can be due to the presence of a texture in the microstructure of the material.

One of the simplest examples is an ideal Langevin superparamagnet, which by definition consists of randomly-oriented noninteracting single-domain nanoparticles (macrospins) that are embedded in a rigid nonmagnetic matrix. The magnetic behavior of this system is determined by the $B/T$~ratio, where $T$ is the absolute temperature. At remanence and not too low $T$, due to the randomizing effect of the thermal energy, the macrospins are randomly oriented in the matrix and the average system magnetization vanishes. The ensuing magnetic SANS cross section is then isotropically distributed in the detector plane. Applying a magnetic field induces an average magnetization, which may result in the appearance of an angular anisotropy of the scattering pattern.

Here, we consider the case of a statistically-isotropic dilute ensemble of identical magnetic nanoparticles, so that case~(iii) is excluded as a source of the scattering anisotropy. Temperature effects are not taken into account. The particles are assumed to be in a single-domain state during the magnetization-reversal process, which implies that the only sample-related angular anisotropy that eventually becomes visible on the detector is due to magnetic anisotropy. As we will see below, cases~(i) and (ii) can be disentangled from one another. The magnetic nanoparticle ensemble is treated within the well-known Stoner-Wohlfarth model~\cite{sw48}, which is a workhorse in magnetism, since it is the simplest approach for producing hysteresis effects. The Stoner-Wohlfarth model considers a system of noninteracting single-domain particles in the presence of an applied magnetic field. The particles exhibit magnetic anisotropy, which may have its origin in the dipolar shape anisotropy and/or in the spin-orbit-interaction-related magnetocrystalline anisotropy. We analyze the role played by the magnetic anisotropy for the angular anisotropy of the two-dimensional magnetic SANS cross section of Stoner-Wohlfarth particles. The Landau-Lifshitz equation of motion for the magnetization is employed to determine the magnetic ground state and to calculate the corresponding magnetic SANS signal and the pair-distance distribution function.

The paper is organized as follows: Section~\ref{msans} displays the well-known equations for the magnetic SANS cross section of a dilute ensemble of uniformly magnetized single-domain particles that are rigidly embedded in a nonmagnetic matrix. We consider the two most-often used scattering geometries, which have the externally applied magnetic field either perpendicular or parallel to the incoming neutron beam. Section~\ref{swmodel} briefly recapitulates the basic expressions of the Stoner-Wohlfarth model, while Section~\ref{results} presents and discusses the results for the SANS observables. We comment on the case of inhomogeneously magnetized particles and on the effect of interparticle correlations (dense packing). Finally, Section~\ref{summary} summarizes the main findings of this work. In the supporting information of this paper, we provide several movies that feature the average magnetization, the two- and one-dimensional magnetic SANS cross section, as well as the pair-distance distribution function, correlation function, and anisotropy parameter during the magnetization-reversal process assuming different magnetic-anisotropy symmetries (compare the six cases in Table~\ref{table1}).

\section{Magnetic SANS cross section of a dilute ensemble of single-domain particles}
\label{msans}

\begin{figure}[tb!]
\centering
\resizebox{0.80\columnwidth}{!}{\includegraphics{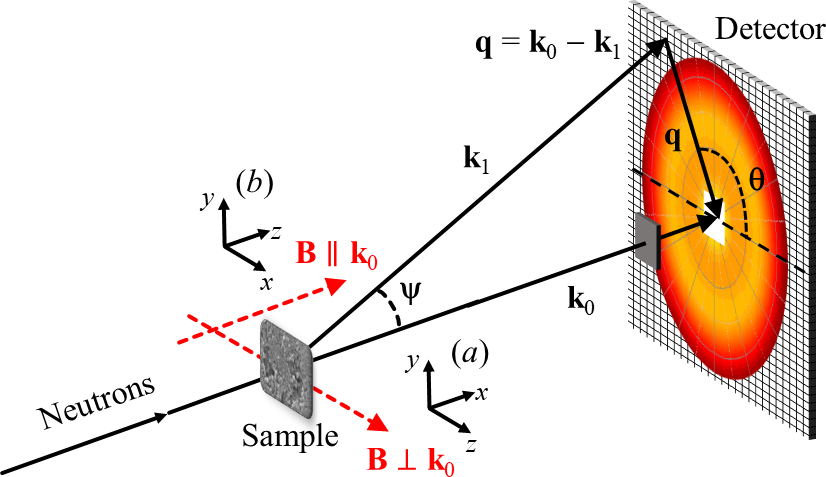}}
\caption{The two most-often employed scattering geometries in magnetic SANS experiments. (\textit{a})~External magnetic field $\mathbf{B}$ perpendicular to the incoming neutron beam; (\textit{b})~$\mathbf{B} \parallel \mathbf{k}_0$. Note that $\mathbf{B} \parallel \mathbf{e}_z$ in both geometries. The momentum-transfer or scattering vector $\mathbf{q}$ corresponds to the difference between the wavevectors of the incident ($\mathbf{k}_0$) and scattered ($\mathbf{k}_1$) neutrons, \textit{i.e.}\ $\mathbf{q} = \mathbf{k}_0 - \mathbf{k}_1$; its magnitude for elastic scattering is given by $q = 4\pi/\lambda \sin(\psi/2)$, where $\lambda$ is the mean wavelength of the neutrons and $\psi$ is the scattering angle. The angle $\theta$ is used in order to describe the angular anisotropy of the recorded scattering pattern on the two-dimensional position-sensitive detector.}
\label{fig1}
\end{figure}

Magnetic SANS experiments are usually conducted with the external magnetic field $\mathbf{B}$ either applied perpendicular ($\perp$) or parallel ($\parallel$) to the wavevector $\mathbf{k}_0$ of the incoming neutron beam (compare Fig.~\ref{fig1}). For these two scattering geometries, the macroscopic elastic magnetic SANS cross section $d\Sigma_M / d\Omega$ at momentum-transfer or scattering vector $\mathbf{q}$ can be expressed as~\cite{michelsbook}:
\begin{align}
\frac{d\Sigma_{M,\perp}^{k}}{d\Omega} = \frac{8\pi^3 b_H^2}{V} \left[ 
\left\langle \left|\widetilde{M}_x^{ik} \right|^2 \right\rangle_i + \left\langle \left|\widetilde{M}_y^{ik} \right|^2 \right\rangle_i \cos^2\theta + \left\langle \left|\widetilde{M}_z^{ik} \right|^2 \right\rangle_i \sin^2\theta \right. \nonumber \\ - \left.
\left\langle \widetilde{M}_y^{ik} \widetilde{M}_z^{\star, ik} + \widetilde{M}_y^{\star, ik} \widetilde{M}_z^{ik} \right\rangle_i \sin\theta \cos\theta \right] 
\label{sigmaperp1} ,
\end{align}
\begin{align}
\frac{d\Sigma_{M,\parallel}^{k}}{d\Omega} = \frac{8\pi^3 b_H^2}{V} \left[ 
\left\langle \left|\widetilde{M}_x^{ik} \right|^2 \right\rangle_i \sin^2\theta + \left\langle \left|\widetilde{M}_y^{ik} \right|^2 \right\rangle_i \cos^2\theta + \left\langle \left|\widetilde{M}_z^{ik} \right|^2 \right\rangle_i \right. \nonumber \\ - \left.
\left\langle \widetilde{M}_x^{ik} \widetilde{M}_y^{\star, ik} + \widetilde{M}_x^{\star, ik} \widetilde{M}_y^{ik} \right\rangle_i \sin\theta \cos\theta \right] 
\label{sigmapara2} .
\end{align}
In these equations, $V$ is the scattering volume, $b_H = 2.91 \times 10^8 \, \mathrm{A}^{-1} \mathrm{m}^{-1}$ denotes the magnetic scattering length in the small-angle regime, $\widetilde{M}_{x,y,z}(\mathbf{q})$ are the Cartesian components of the magnetization vector field $\mathbf{M}(\mathbf{r}) = [ M_x(\mathbf{r}), M_y(\mathbf{r}), M_z(\mathbf{r}) ]$, the index $i$ refers to the orientation of particle $i$, the index $k$ keeps track of the applied-field value, the asterisk $\star$ denotes the complex-conjugated quantity, and the $\left\langle ... \right\rangle_i$ notation is explained below. In the perpendicular and parallel scattering geometries, the scattering vector is given by $\mathbf{q}_{\perp} = q [ 0 , \sin\theta, \cos\theta ]$ and $\mathbf{q}_{\parallel} = q [ \cos\theta, \sin\theta, 0 ]$, where the angle $\theta$ is, respectively, measured between $\mathbf{q}_{\perp}$ and $\mathbf{B} \parallel \mathbf{e}_z$ and $\mathbf{q}_{\parallel}$ and $\mathbf{e}_x$. Note that $\mathbf{B} \parallel \mathbf{e}_z$ in both geometries. Equations~(\ref{sigmaperp1}) and (\ref{sigmapara2}) neglect interparticle interference effects and are valid for a dilute scattering system.

In the main part of the paper we exclusively focus on the perpendicular scattering geometry, and we refer to the supporting information for videos that show results for the parallel geometry.

The Fourier transform of the magnetization vector field of a nanoparticle is defined by
\begin{align}
\widetilde{\mathbf{M}}(\mathbf{q}) = \frac{1}{(2\pi)^{3/2}} \int_{V_p} \mathbf{M}(\mathbf{r}) \exp\left( - \mathrm{i}\mathbf{q} \cdot \mathbf{r}\right) d^3r ,
\label{fourier1}
\end{align}
which for a uniformly magnetized particle can be simplified to
\begin{align}
\widetilde{\mathbf{M}}(\mathbf{q}) = \frac{M_0 \mathbf{m}}{(2\pi)^{3/2}} \int_{V_p} \exp\left( - \mathrm{i}\mathbf{q} \cdot \mathbf{r}\right) d^3r .
\label{fourier2}
\end{align}
In going from equation~(\ref{fourier1}) to equation~(\ref{fourier2}), we have expressed the (constant) magnetization vector as $\mathbf{M} = M_0 \mathbf{m}$, where $M_0$ is the saturation magnetization and $\mathbf{m}$ is a unit vector along $\mathbf{M}$. The remaining integral in equation~(\ref{fourier2}) is the well-known form-factor integral (over the volume $V_p$ of the particle), which is analytically known for many particle shapes. For spherical particles (with radius $R$), equation~(\ref{fourier2}) can be further simplified to:
\begin{align}
\widetilde{\mathbf{M}}(\mathbf{q}) = \frac{4\pi R^3 M_0}{(2\pi)^{3/2}} \frac{j_1(q R)}{q R} \mathbf{m} ,
\label{fourier3}
\end{align}
where $j_1(z)$ denotes the first-order spherical Bessel function. For more complicated particle shapes (\textit{e.g.}\ cylinders or flat discs), an additional average over the particle orientation might be required to obtain the only $q$~dependent form factor. To remind the reader, we consider a dilute system of $N$ identical spherical single-domain particles, where, at a given value of the applied field $B_k$, each particle $i$ has its own random orientation of magnetic easy axes with respect to $\mathbf{B}$ (to be further specified in Section~\ref{swmodel}). For this situation, we may express the Fourier components as
\begin{align}
\widetilde{\mathbf{M}}^{ik}(\mathbf{q}) = \frac{4\pi R^3 M_0}{(2\pi)^{3/2}} \frac{j_1(q R)}{q R} \mathbf{m}^{ik} . 
\end{align}
The magnetic SANS cross sections at the $k$-th magnetic field value, averaged over all the random easy-axis orientations $i$, is then given by equations~(\ref{sigmaperp1}) and (\ref{sigmapara2}), where the bracket notation of the mean operator is defined as follows:
\begin{align}
\left\langle \left|\widetilde{M}_x^{i,k} \right|^2 \right\rangle_i = \frac{1}{N} \sum_{i=1}^{N} \left|\widetilde{M}_x^{i,k} \right|^2 ,
\end{align}
and similar for the other Fourier components. Since the spherical Bessel function is a scalar prefactor to the Fourier transform of the magnetization vector, we can simplify the magnetic SANS cross sections as follows: 
\begin{align}
\frac{d\Sigma_{M,\perp}^{k}}{d\Omega} = \frac{16\pi^2 R^6 M_0^2 b_H^2}{V} \left(\frac{j_1(q R)}{q R}\right)^2 \nonumber \\
\times \left[ \left\langle (m_x^{ik})^2 \right\rangle_i + \left\langle (m_y^{ik})^2 \right\rangle_i \cos^2\theta + \left\langle (m_z^{ik})^2 \right\rangle_i \sin^2\theta - 2 \left\langle m_y^{ik} m_z^{ik} \right\rangle_i \sin\theta \cos\theta \right] ,
\end{align}
\begin{align}
\frac{d\Sigma_{M,\parallel}^{k}}{d\Omega} = \frac{16\pi^2 R^6 M_0^2 b_H^2}{V} \left(\frac{j_1(q R)}{q R}\right)^2 \nonumber \\
\times \left[ \left\langle (m_x^{ik})^2 \right\rangle_i \sin^2\theta + \left\langle (m_y^{ik})^2 \right\rangle_i \cos^2\theta + \left\langle (m_z^{ik})^2 \right\rangle_i - 2 \left\langle m_x^{ik} m_y^{ik} \right\rangle_i \sin\theta \cos\theta \right] ,
\end{align}
such that the magnetic SANS cross section directly follows from the (real-valued) real-space correlation functions of the components of the magnetization vector. We therefore define the crosscorrelation matrix corresponding to the $k$-th value of the magnetic field as:
\begin{align}
\boldsymbol{\Gamma}^{k} = 
\left\langle \mathbf{m}^{ik} \otimes \mathbf{m}^{ik} \right\rangle_i ,
\end{align}
where more explicitly written the components are defined as:
\begin{align}
\Gamma_{\alpha\beta}^{k} = \frac{1}{N} \sum_{i=1}^{N} m_{\alpha}^{i,k}m_{\beta}^{i,k} , \quad \alpha,\beta\in\{x,y,z\}
\label{crosscorrdef}
\end{align}
such that the magnetic SANS cross sections are written as:
\begin{align}
\frac{d\Sigma_{M,\perp}^{k}}{d\Omega} = \frac{16\pi^2 R^6 M_0^2 b_H^2}{V} \left(\frac{j_1(q R)}{q R}\right)^2 \nonumber \\
\times \left[ \Gamma_{xx}^{k} +\Gamma_{yy}^{k} \cos^2\theta + \Gamma_{zz}^{k} \sin^2\theta - 2 \Gamma_{yz}^{k} \sin\theta \cos\theta \right] ,
\label{sigmaperpxy}
\end{align}
\begin{align}
\frac{d\Sigma_{M,\parallel}^{k}}{d\Omega} = \frac{16\pi^2 R^6 M_0^2 b_H^2}{V} \left(\frac{j_1(q R)}{q R}\right)^2 \nonumber \\
\times \left[ \Gamma_{xx}^{k} \sin^2\theta + \Gamma_{yy}^{k} \cos^2\theta + \Gamma_{zz}^{k} - 2 \Gamma_{xy}^{k} \sin\theta \cos\theta \right] .
\label{sigmaparaxy}
\end{align}
One may note that in the general case of an inhomogeneous magnetization distribution, \textit{i.e.}\ $\mathbf{m} = \mathbf{m}(\mathbf{r})$, these formulations of the magnetic SANS cross sections only correspond to the first term in a Taylor series expansion. From equations~(\ref{sigmaperpxy}) and (\ref{sigmaparaxy}) we see that the magnetic SANS cross section gives quite different insights into the magnetization structure than the hysteresis loop: the latter contains information about the first-order moments, while the former yields information about the second-order moments of the magnetization vector field. The magnetic SANS cross section of uniformly magnetized particles is anisotropic ($\theta$~dependent) when the terms in the square brackets on the second lines of equations~(\ref{sigmaperpxy}) and (\ref{sigmaparaxy}) add up to yield a resulting net $\theta$~dependence. This statement is true for arbitrary particle shape and also in the presence of a distribution of particle sizes, as long as all the particles are in a single-domain state.

The azimuthally-averaged SANS cross sections are straightforwardly obtained as: 
\begin{align}
I_{\perp}^{k}(q) &= \frac{1}{2\pi} \int_{0}^{2\pi} \frac{d\Sigma_{M,\perp}^{k}}{d\Omega} \; d \theta \nonumber \\ &= \frac{16\pi^2R^6M_0^2b_H^2}{V}\left(\frac{j_1(q R)}{q R}\right)^2 \left[ \Gamma_{xx}^{k} + 0.5 \Gamma_{yy}^{k} + 0.5 \Gamma_{zz}^{k} \right] ,
\label{aziperp}
\end{align}
\begin{align}
I_{\parallel}^{k}(q) &= \frac{1}{2\pi} \int_{0}^{2\pi} \frac{d\Sigma_{M,\parallel}^{k}}{d\Omega} \; d \theta \nonumber \\ &= \frac{16\pi^2R^6M_0^2b_H^2}{V}\left(\frac{j_1(q R)}{q R}\right)^2 \left[ 0.5 \Gamma_{xx}^{k} + 0.5 \Gamma_{yy}^{k} + \Gamma_{zz}^{k} \right] .
\label{azipara}
\end{align}
Here, for the $2\pi$~azimuthal average, we see that the $yz$ and $xy$~crosscorrelation terms vanish and only the autocorrelation terms remain. The pair-distance distribution functions are obtained as:
\begin{align}
p_{\perp}^{k}(r) &= r^2 \int_{0}^{\infty} I_{\perp}^{k}(q) j_0(q r) q^2 dq \nonumber \\ &=\frac{16\pi^2 R^3 M_0^2 b_H^2}{V} \frac{\pi}{6} r^2 \left(1 - \frac{3r}{4R} + \frac{r^3}{16R^3} \right) \left[ \Gamma_{xx}^{k} + 0.5 \Gamma_{yy}^{k} + 0.5 \Gamma_{zz}^{k} \right] ,
\label{pvonrperp}
\end{align}
\begin{align}
p_{\parallel}^{k}(r) &= r^2 \int_{0}^{\infty} I_{\parallel}^{k}(q) j_0(q r) q^2 dq \nonumber \\ &=\frac{16\pi^2 R^3 M_0^2 b_H^2}{V} \frac{\pi}{6} r^2 \left(1 - \frac{3r}{4R} + \frac{r^3}{16R^3} \right) \left[ 0.5 \Gamma_{xx}^{k} + 0.5 \Gamma_{yy}^{k} + \Gamma_{zz}^{k} \right] .
\label{pvonrpara}
\end{align}
The related correlation functions $c^{k}(r) = p^{k}(r)/r^2$ are:
\begin{align}
c_{\perp}^{k}(r) = \frac{16\pi^2 R^3 M_0^2 b_H^2}{V} \frac{\pi}{6} \left(1 - \frac{3r}{4R} + \frac{r^3}{16R^3} \right) \left[ \Gamma_{xx}^{k} + 0.5 \Gamma_{yy}^{k} + 0.5 \Gamma_{zz}^{k} \right] ,
\label{cvonrperp}
\end{align}
\begin{align}
c_{\parallel}^{k}(r) = \frac{16\pi^2 R^3 M_0^2 b_H^2}{V} \frac{\pi}{6} \left(1 - \frac{3r}{4R} + \frac{r^3}{16R^3} \right) \left[ 0.5 \Gamma_{xx}^{k} + 0.5 \Gamma_{yy}^{k} + \Gamma_{zz}^{k} \right] .
\label{cvonrpara}
\end{align}
The $\Gamma_{\alpha\beta}^{k}$ are real numbers, which depend on the applied field and on the symmetry of the magnetic anisotropy of the particles. We then see that---within the present Stoner-Wohlfarth approach---the $I(q)$, $p(r)$, and $c(r)$ are identical for the perpendicular and parallel scattering geometries, except for a numerical prefactor.

To quantify the angular anisotropy of the two-dimensional magnetic SANS cross section, we introduce (for $\mathbf{B} \perp \mathbf{k}_0$) the following number [compare Fig.~\ref{fig1} and equation~(\ref{sigmaperpxy})]:
\begin{align}
A_{\perp}^{k} = \frac{\int_{0}^{\infty} \left. \frac{d\Sigma_{M,\perp}^{k}}{d\Omega} \right|_{\mathbf{q} \parallel \mathbf{e}_z} dq}{\int_{0}^{\infty} \left. \frac{d\Sigma_{M,\perp}^{k}}{d\Omega} \right|_{\mathbf{q} \parallel \mathbf{e}_y} dq} = \frac{\Gamma_{xx}^k + \Gamma_{yy}^k}{\Gamma_{xx}^k + \Gamma_{zz}^k} .
\label{aninumber1}
\end{align}
For the parallel scattering geometry, where $\mathbf{B}$ is perpendicular to the detector plane, we calculate $A^{k}$ similar to equation~(\ref{aninumber1}) as the ratio of integrated intensities along the horizontal and vertical directions on the detector [compare equation~(\ref{sigmaparaxy})]:
\begin{align}
A_{\parallel}^{k} = \frac{\int_{0}^{\infty} \left. \frac{d\Sigma_{M,\parallel}^{k}}{d\Omega} \right|_{\mathbf{q} \parallel \mathbf{e}_x} dq}{\int_{0}^{\infty} \left. \frac{d\Sigma_{M,\parallel}^{k}}{d\Omega} \right|_{\mathbf{q} \parallel \mathbf{e}_y} dq} = \frac{\Gamma_{yy}^k + \Gamma_{zz}^k}{\Gamma_{xx}^k + \Gamma_{zz}^k} .
\label{aninumber2}
\end{align}

\section{Recap: Stoner-Wohlfarth model}
\label{swmodel}

In this chapter, we recapitulate the basic ideas of the Stoner-Wohlfarth model~\cite{sw48}, which considers a magnetically-anisotropic single-domain particle in the presence of an applied magnetic field $\mathbf{B}$ (assumed here to be parallel to the $z$~direction of a Cartesian laboratory coordinate system). The origin of the magnetic anisotropy can be due to shape anisotropy and/or magnetocrystalline anisotropy. Here, we consider identical particles possessing magnetocrystalline anisotropy only. Note also that spherical particles do not exhibit shape anisotropy. Thermal effects are ignored. Denoting with $\omega_{\mathrm{ani}}^{ik}$ the magnetic anisotropy energy density in the (global) laboratory coordinate system, the total energy density $\omega^{ik}$ of a particle $i$ at field $k$ is commonly expressed as: 
\begin{align}
\omega^{ik}(\mathbf{m}^{ik}) = - M_0 \mathbf{m}^{ik} \cdot \mathbf{B}^k + \omega_{\mathrm{ani}}^{ik}(\mathbf{m}^{ik}) ,
\label{edensity}
\end{align}
where $M_0$ is the saturation magnetization of the material. The two most-common forms of magnetocrystalline anisotropy either exhibit uniaxial (u) or cubic (c) symmetry. The corresponding mathematical expressions for the magnetic anisotropy energy densities, in the local coordinate frame of the particle, are the following:
\begin{align}
\omega_{\mathrm{ani, u}}(\mathbf{m})  = K_{\mathrm{u}1} \left( 1 - m_z^2 \right) + K_{\mathrm{u}2} \left( 1 - m_z^2 \right)^2 ,
\label{eq:UniaxialAnisotropyEnergyDensity}
\end{align}
\begin{align}
\omega_{\mathrm{ani, c}}(\mathbf{m})  = K_{\mathrm{c}1} \left( m_x^2 m_y^2 + m_x^2 m_z^2 + m_y^2 m_z^2 \right) + K_{\mathrm{c}2} m_x^2 m_y^2 m_z^2 ,
\label{eq:CubicAnisotropyEnergyDensity}
\end{align}
where the $K_{\mathrm{u}}$ and $K_{\mathrm{c}}$ are the temperature-dependent anisotropy constants (in $\mathrm{J}/\mathrm{m^3}$). Depending on their relative magnitude and the signs of the anisotropy constants, different easy axes are obtained~\cite{kronfahn03}. The corresponding effective magnetic fields (in Tesla) are then readily obtained as:
\begin{align}
\mathbf{B}_{\mathrm{ani,u}}(\mathbf{m}) = - \frac{1}{M_0} \frac{\partial \omega_{\mathrm{ani, u}}}{\partial \mathbf{m}} = \frac{2 K_{\mathrm{u}1}}{M_0} 
\begin{bmatrix}
0 \\
0 \\
m_z
\end{bmatrix} + \frac{4 K_{\mathrm{u}2}}{M_0}
\begin{bmatrix}
0 \\
0 \\
m_z \left( 1 - m_z^2 \right)
\end{bmatrix} ,
\label{evelyn1}
\end{align}
\begin{align}
\mathbf{B}_{\mathrm{ani,c}}(\mathbf{m}) = - \frac{1}{M_0} \frac{\partial \omega_{\mathrm{ani, c}}}{\partial \mathbf{m}} = - \frac{2 K_{\mathrm{c}1}}{M_0} 
\begin{bmatrix}
m_x \left( m_y^2 + m_z^2 \right) \\
m_y \left( m_x^2 + m_z^2 \right) \\
m_z \left( m_x^2 + m_y^2 \right)
\end{bmatrix} - \frac{2 K_{\mathrm{c}2}}{M_0}
\begin{bmatrix}
m_x m_y^2 m_z^2 \\
m_x^2 m_y m_z^2 \\
m_x^2 m_y^2 m_z
\end{bmatrix} .
\label{evelyn2}
\end{align}
The different (random) particle orientations $i$ are obtained by rotations in three-dimensional space (change of basis). This is accomplished by using a rotation matrix $\mathbf{R}_i$ that is parametrized by (random) Euler angles $\gamma_i, \delta_i, \epsilon_i$ (with $0 \le \gamma_i \le 2\pi$, $0 \le \delta_i \le \pi$, $0 \le \epsilon_i \le 2\pi$)~\cite{goldstein}, so that the total effective field is calculated as follows:
\begin{align}
\mathbf{B}_{\mathrm{eff}}^{ik}(\mathbf{m}^{ik}) = - \frac{1}{M_0} \frac{\partial \omega^{ik}}{\partial \mathbf{m}^{ik}} = \mathbf{B}^k + \mathbf{R}_i \cdot \mathbf{B}_{\mathrm{ani}}(\mathbf{R}_i^{\mathrm{T}} \cdot \mathbf{m}^{ik}) ,
\label{eq:beff}
\end{align}
where
\begin{align}
\textsl{\textbf{R}}_i(\gamma_i, \delta_i, \epsilon_i) = \textsl{\textbf{R}}_z(\gamma_i) \cdot \textsl{\textbf{R}}_y(\delta_i) \cdot \textsl{\textbf{R}}_z(\epsilon_i) ,
\end{align}
and the superscript ${\mathrm{T}}$ refers to the transpose of the matrix. Note that we adopt here a $z$$-$$y$$-$$z$~rotation sequence. The procedure of obtaining the $\gamma_i, \delta_i, \epsilon_i$ starts with uniformly distributed random numbers $a_i, b_i, c_i$ in the three-dimensional unit cube, such that $0 \le a_i, b_i, c_i \le 1$. As a random number generator we use the low-discrepancy Sobol sequences~\cite{sobol}. In order to achieve a uniform distribution of random angles on the unit sphere, we use the following transformations: 
\begin{align}
\gamma_i &= 2\pi a_i , \\
\delta_i &= \arccos(2 b_i - 1), \\
\epsilon_i &= 2\pi c_i .
\end{align}
To obtain the static equilibrium magnetization, we insert the expression for $\mathbf{B}_{\mathrm{eff}}^{ik}$ into the Landau-Lifshitz (LL) equation~\cite{bertottibook}, which describes the magnetization dynamics:
\begin{align}
\frac{d\mathbf{m}^{ik}}{dt} = - \gamma_{\mathrm{G}} \, \mathbf{m}^{ik} \times \mathbf{B}_{\mathrm{eff}}^{ik} - \eta \, \mathbf{m}^{ik}\times (\mathbf{m}^{ik}\times\mathbf{B}_{\mathrm{eff}}^{ik}) ,
\end{align}
where $\gamma_{\mathrm{G}} = 1.76 \times 10^{11} \, \mathrm{T}^{-1} \mathrm{s}^{-1}$ is the gyromagnetic ratio and $\eta$ the damping constant. Following the temporal evolution of the LL equation, the static spin structure $\mathbf{m}^{ik} = [m_x^{ik}, m_y^{ik}, m_z^{ik} ]$ of nanomagnet $i$ at field $k$ can be obtained. Repeating (at fixed $k$) these simulations $N$~times for different easy-axis orientations allows us to compute the averages that determine the magnetic SANS cross section. More specifically, the hysteresis loop of the ensemble of spherical nanomagnets then follows from the averaged magnetization projected along the $z$~direction, 
\begin{align}
\overline{m}_z^{k} = \left\langle m_z^{i,k} \right\rangle_i = \frac{1}{N}\sum_{i=1}^{N} m_z^{i,k} .
\end{align}
In addition to $\overline{m}_z^{k}$ we also calculate the field dependence of the transversal magnetization components, $\overline{m}_x^{k}$ and $\overline{m}_y^{k}$, as well as the field loops of the components of the crosscorrelation matrix $\Gamma_{\alpha,\beta}$ (with $\alpha,\beta \in\{x,y,z\}$) [equation~(\ref{crosscorrdef})]. As we have seen in Section~\ref{msans}, these are of particular relevance for the magnetic SANS cross section. For further details on the SANS simulation methodology using the LL equation, we refer to Ref.~\cite{adamsjacnum2022}.

In the numerical computations, we used the following parameters: $\eta = 3 \times 10^{11} \, \mathrm{T}^{-1} \mathrm{s}^{-1}$, and an integration time step of $5 \times 10^{-15} \, \mathrm{s}$. Typically $K = 2000$ discretization points for the applied magnetic field and $N = 10000$ samples of different orientations for the easy axes of the particles (angles $\gamma_i, \delta_i, \epsilon_i$) were used.

\section{Results and Discussion}
\label{results}

In our analysis, we consider the following six cases for Stoner-Wohlfarth particles with uniaxial and cubic anisotropy: (i)~$k_{\mathrm{u}1} = +1$ and $k_{\mathrm{u}2} = 0$, (ii)~$k_{\mathrm{u}1} = -1$ and $k_{\mathrm{u}2} = 0$, (iii)~$k_{\mathrm{u}1} = -0.5$ and $k_{\mathrm{u}2} = +0.5$, (iv)~$k_{\mathrm{c}1} = +1$ and $k_{\mathrm{c}2} = 0$, (v)~$k_{\mathrm{c}1} = -1$ and $k_{\mathrm{c}2} = 0$, (vi)~$k_{\mathrm{c}1} = -1$ and $k_{\mathrm{c}2} = +9$. Note that the $k_{\mathrm{u,c}}$ (in Tesla) are related to the $K_{\mathrm{u,c}}$ (in $\mathrm{J}/\mathrm{m^3}$) via $k_{\mathrm{u,c}} = 2 K_{\mathrm{u,c}}/M_0$ [compare equations~(\ref{evelyn1}) and (\ref{evelyn2})]. Minimization of the anisotropy energy densities [equations~(\ref{eq:UniaxialAnisotropyEnergyDensity}) and (\ref{eq:CubicAnisotropyEnergyDensity})] shows that these combinations of anisotropy constants correspond to the following well-known easy-axis orientations in hexagonal and cubic single crystals~\cite{kronfahn03}: (i)~easy $c$-axis, (ii)~easy basal plane, (iii)~easy cone with opening angle $\sin\phi = \sqrt{-k_{\mathrm{u}1}/(2 k_{\mathrm{u}2})}$, (iv)~$\langle 100 \rangle$ directions, (v)~$\langle 111 \rangle$ directions, (vi)~$\langle 110 \rangle$ directions. We refer to the supporting information of this paper, where several movies that feature the average magnetization and the SANS observables during the magnetization-reversal process are provided.

\begin{figure}[tb!]
\centering
\resizebox{1.0\columnwidth}{!}{\includegraphics{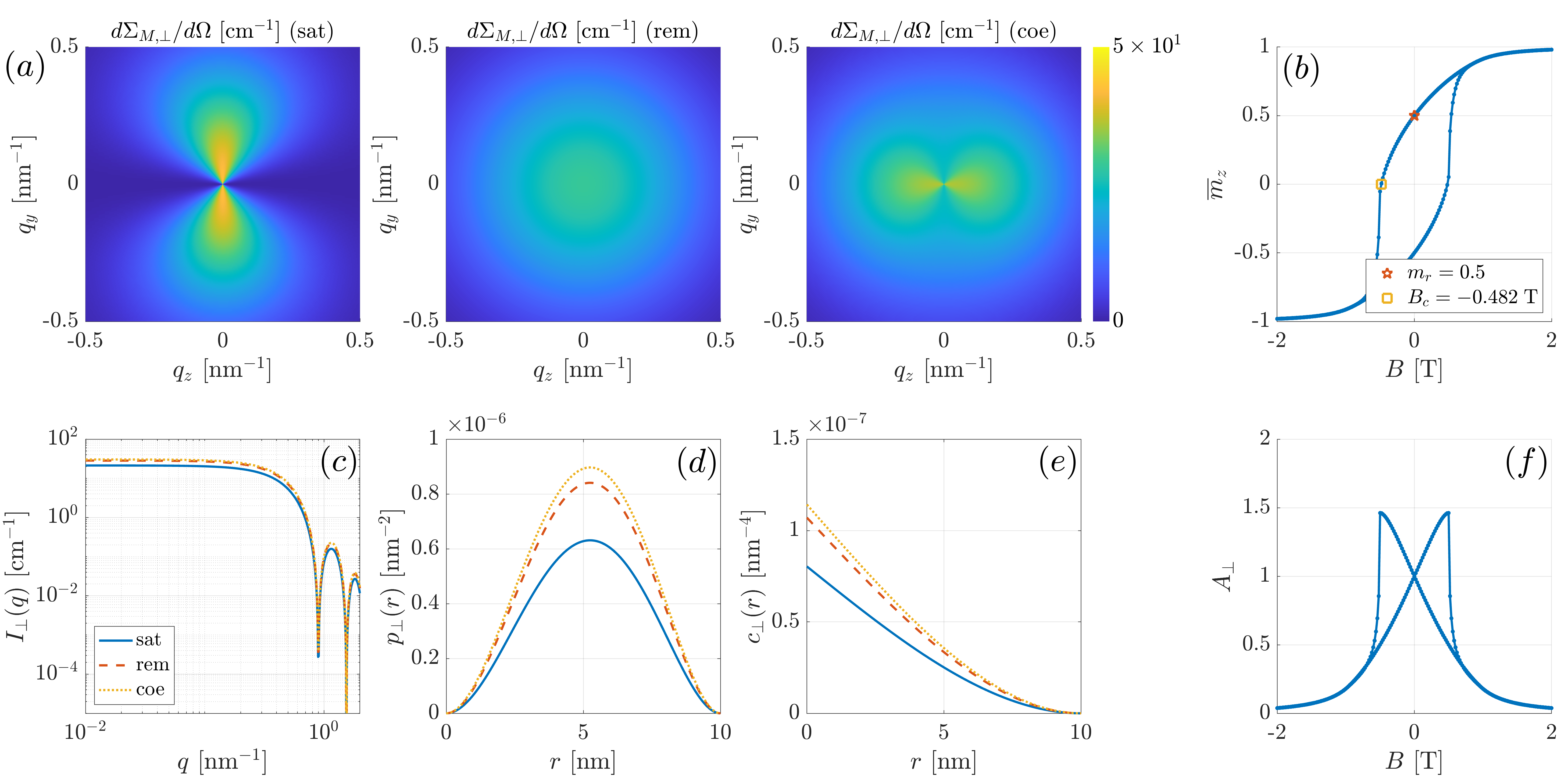}}
\caption{Results for the magnetization and the SANS observables of a dilute ensemble of uniaxial Stoner-Wohlfarth particles with $k_{\mathrm{u}1} = +1$ and $k_{\mathrm{u}2} = 0$ [case~(i)] ($\mathbf{B} \perp \mathbf{k}_0$). The (spherical) particle diameter is $D = 10 \, \mathrm{nm}$. (\textit{a})~Two-dimensional magnetic SANS cross sections $d\Sigma_{M,\perp} / d\Omega$ at saturation (sat), remanence (rem), and at the coercive field (coe); (\textit{b})~hysteresis loop $\overline{m}_z(B)$ (the reduced remanence and the coercivity are indicated); (\textit{c})~one-dimensional $2\pi$~azimuthally-averaged magnetic SANS cross sections $I_{\perp}(q) = \frac{d\Sigma_{M,\perp}}{d\Omega}(q)$; (\textit{d})~pair-distance distribution functions $p_{\perp}(r)$; (\textit{e}) correlation functions $c_{\perp}(r)$; (\textit{f})~anisotropy parameter $A_{\perp}(B)$.}
\label{fig2}
\end{figure}

Table~\ref{table1} contains the values of the second moments $\Gamma_{\alpha \alpha}$ of the components of the magnetization vectors at selected points on the hysteresis loop (remanence and coercivity) and for the different anisotropy symmetries. The hysteresis loops and the full field dependencies of the autocorrelations are shown in Appendix~\ref{appendixA}. As an example, for uniaxial Stoner-Wohlfarth particles of case~(i), Fig.~\ref{fig2} depicts the results for the SANS observables. Inspection of the table entries for the $\Gamma_{\alpha \alpha}$ and comparison to the magnetic SANS cross sections [equations~(\ref{sigmaperpxy}) and (\ref{sigmaparaxy})] reveals that only case~(i) yields, in the perpendicular scattering geometry, an isotropic two-dimensional SANS image at remanence. In all other cases do we find (for $\mathbf{B} \perp \mathbf{k}_0$) an anisotropic magnetic SANS pattern at remanence and at the coercive field. By contrast, in the parallel scattering geometry ($\mathbf{B} \parallel \mathbf{k}_0$), we observe (since $\Gamma_{xx} = \Gamma_{yy}$) an isotropic magnetic $d\Sigma_{M,\parallel} / d\Omega$ at all fields during the magnetization-reversal process.

\begin{table}[tb!]
\caption{\label{table1}
Stoner-Wohlfarth particles with uniaxial and cubic anisotropy~\cite{usov1997}:~Values for the reduced remanence $m_r$, the coercivity $B_c$ (in Tesla), and for the autocorrelations $\Gamma_{xx}$, $\Gamma_{yy}$, and $\Gamma_{zz}$ [equation~(\ref{crosscorrdef})] at these points on the hysteresis loop. All crosscorrelations $\Gamma_{\alpha \beta}$ with $\alpha \neq \beta$ vanish. The $k_{\mathrm{u,c}}$ are given in units of Tesla.}
\begin{ruledtabular}
\begin{tabular}{lccccccccc}
 & $m_r$ & $B_c$ & $\Gamma_{xx}^{m_r}$ & $\Gamma_{yy}^{m_r}$ & $\Gamma_{zz}^{m_r}$ & $\Gamma_{xx}^{B_c}$ & $\Gamma_{yy}^{B_c}$ & $\Gamma_{zz}^{B_c}$ \\
\hline
case~(i): uniaxial \\ ($k_{\mathrm{u}1} = +1$, $k_{\mathrm{u}2} = 0$) & 0.5 & 0.482 & 0.333 & 0.333 & 0.333 & 0.422 & 0.422 & 0.156 \\
case~(ii): uniaxial \\ ($k_{\mathrm{u}1} = -1$, $k_{\mathrm{u}2} = 0$) & 0.785 & 0 & 0.167 & 0.167 & 0.667 & --- & --- & --- \\
case~(iii): uniaxial \\ ($k_{\mathrm{u}1} = -0.5$, $k_{\mathrm{u}2} = +0.5$) & 0.909 & 0 & 0.083 & 0.083 & 0.833 & --- & --- & --- \\
case~(iv): cubic \\ ($k_{\mathrm{c}1} = +1$, $k_{\mathrm{c}2} = 0$) & 0.831 & 0.321 & 0.150 & 0.150 & 0.700 & 0.1875 & 0.1875 & 0.625 \\
case~(v): cubic \\ ($k_{\mathrm{c}1} = -1$, $k_{\mathrm{c}2} = 0$) & 0.866 & 0.189 & 0.121 & 0.121 & 0.758 & 0.225 & 0.225 & 0.550 \\
case~(vi): cubic \\ ($k_{\mathrm{c}1} = -1$, $k_{\mathrm{c}2} = +9$) &0.912 & 0.383 & 0.082 & 0.082 & 0.836 & 0.105 & 0.105 & 0.790 \\
\end{tabular}
\end{ruledtabular}
\end{table}

Considering case~(i), we see that the angular anisotropy of the two-dimensional $d\Sigma_{M,\perp} / d\Omega$ changes strongly between saturation ($\sin^2\theta$~type), remanence (isotropic), and the coercive field ($\cos^2\theta$~type) [Fig.~\ref{fig2}(\textit{a})], while the azimuthally-averaged $d\Sigma_{M,\perp} / d\Omega$ change relatively little between these situations [Fig.~\ref{fig2}(\textit{c})]. Decreasing the field from saturation (where $\Gamma_{xx} = \Gamma_{yy} = 0$ and $\Gamma_{zz} = 1$) to zero field and to $B_c$, we observe an increase of the one-dimensional $d\Sigma_{M,\perp} / d\Omega$ and of the pair-distance distribution function $p_{\perp}(r)$ [Fig.~\ref{fig2}(\textit{d})] and the correlation function $c_{\perp}(r)$ [Fig.~\ref{fig2}(\textit{e})] [compare equations~(\ref{pvonrperp}) to (\ref{cvonrpara})]. From these results it may be concluded that, depending on the anisotropy symmetry of Stoner-Wohlfarth particles, an anisotropic magnetic SANS pattern is (generally) obtained; compare Fig.~\ref{fig2}(\textit{f}) for the anisotropy parameter $A_{\perp}(B)$. In experimental studies, where often the two-dimensional \textit{total} (nuclear and magnetic) $d\Sigma_{\perp} / d\Omega$ is analyzed, one should therefore be cautious assuming that an isotropic pattern is to be expected at characteristic field values such as at remanence or at the coercive field. To access the purely magnetic SANS cross section in unpolarized experiments, the subtraction of the total $d\Sigma_{\perp} / d\Omega$  at a field close to magnetic saturation from the data at lower fields might help. As shown in Ref.~\cite{bersweiler2019} on Mn-Zn ferrite nanoparticles, an isotropic total $d\Sigma_{\perp} / d\Omega$ at zero field can then turn into an anisotropic purely magnetic signal, in this way providing access to the magnetic correlations. Of course, polarization analysis also yields the purely magnetic SANS cross section, albeit with much more effort regarding the experimental setup and the data-reduction procedure.

As mentioned already in Section~\ref{msans} when discussing equations~(\ref{sigmaperpxy}) and (\ref{sigmaparaxy}), within the present Stoner-Wohlfarth approach, the presence of a particle-size distribution results in the smearing of the form-factor oscillations (\textit{i.e.}\ affects the $q$~dependence), but leaves the angular anisotopy of $d\Sigma_{M,\perp}/ d\Omega$ unchanged.

So far the discussion is based on uniformly magnetized Stoner-Wohlfarth particles. For nonuniformly magnetized nanoparticles, where the magnetization vector field $\mathbf{m} = \mathbf{m}(\mathbf{r})$ is a function of the position $\mathbf{r}$ within the particle, the $d\Sigma_{M,\perp} / d\Omega$ at remanence or at the coercive field is generally also to be expected to depend on the angle $\theta$ in the detector plane. Nonuniformities in the magnetization distribution of nanoparticles are \textit{e.g.}\ caused by surface anisotropy, vacancies, or antiphase boundaries~\cite{nedelkoski2017,krycka2019,zakutna2020,lakbenderdisch2021,koehlerjac2021,dirkreview2022,adamsjacana2022,adamsjacnum2022,evelynprb2022}. Micromagnetic simulations that take into account the relevant interactions such as isotropic exchange, antisymmetric exchange, magnetic anisotropy, Zeeman energy, and the magnetodipolar interaction are an important tool for advancing the understanding of magnetic SANS of nanomagnets~\cite{michelsbook}. Unfortunately, due to the nonlinearity of the underlying integro-differential equations of micromagnetics, numerical simulations have to be carried out.

\begin{figure}[tb!]
\centering
\resizebox{0.90\columnwidth}{!}{\includegraphics{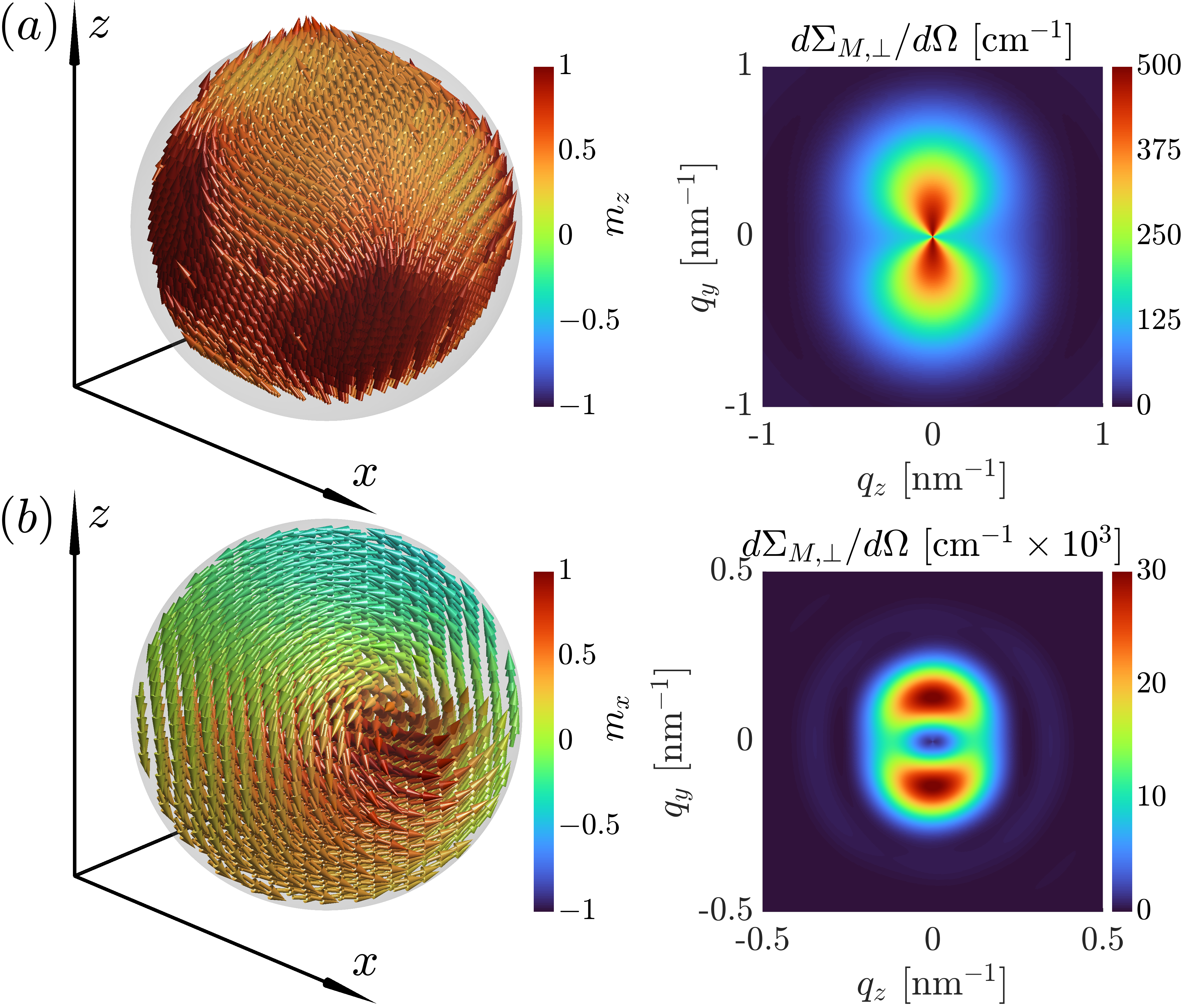}}
\caption{Micromagnetic simulation results for dilute ensembles of inhomogeneously magnetized nanoparticles. (\textit{a})~Remanent spin structure of a spherical $8 \, \mathrm{nm}$-sized nanoparticle with strong N$\mathrm{\acute{e}}$el surface anisotropy and corresponding randomly-averaged magnetic SANS cross section $d\Sigma_{M,\perp} / d\Omega$ ($\mathbf{B} \perp \mathbf{k}_0$)~\cite{adamsjacnum2022}. (\textit{b})~Remanent spin structure and randomly-averaged $d\Sigma_{M,\perp} / d\Omega$ of spherical nanoparticles with a diameter of $32 \, \mathrm{nm}$~\cite{evelynprb2022}.}
\label{fig3}
\end{figure}

As an example, we show in Fig.~\ref{fig3} selected results that feature an anisotropic (randomly-averaged) $d\Sigma_{M,\perp} / d\Omega$ at remanence; for details on the micromagnetic SANS simulation methodology, see Refs.~\cite{adamsjacnum2022,evelynprb2022}. Fig.~\ref{fig3}(\textit{a}) showcases the results of atomistic SANS simulations, where the focus is set on the effect of the N$\mathrm{\acute{e}}$el surface anisotropy on the spin structure and ensuing magnetic SANS signal of randomly-oriented nanoparticles. The snapshot of the real-space spin structure clearly reveals a significant spin disorder in the near-surface region of the nanoparticle, with a corresponding characteristic $\sin^2\theta$~type anisotropic magnetic SANS pattern. Fig.~\ref{fig3}(\textit{b}) displays the results for a random ensemble of spherical nanoparticles. In this system (with no surface anisotropy), the magnetization distribution is determined by the dipolar interaction energy, which gives rise to a vortex-type spin texture. The randomly-averaged $d\Sigma_{M,\perp} / d\Omega$ also exhibits a pronounced $\theta$~dependence.

The simulation results in Fig.~\ref{fig3} were obtained for a dilute set of nanoparticles, \textit{i.e.}\ interparticle correlations are not taken into account. When the particle concentration in a sample increases \textit{positional correlations} become important, which is taken into account by the structure factor $S(q) = \langle \sum_{i,j} \exp(- i \mathbf{q} \cdot \mathbf{r}_{ij} ) \rangle$, where $\mathbf{r}_{ij} = \mathbf{r}_j - \mathbf{r}_i$ denotes the vector connecting the position vectors of particle $i$ and $j$, and the bracket $\langle ... \rangle$ refers to an orientational average. For magnetic particles, whether uniformly or nonuniformly magnetized, additional \textit{magnetic-moment correlations} become relevant, resulting in the appearance of a magnetic structure factor. This has been realized by Honecker~\textit{et al.}~\cite{honecker2020}, who have shown that the magnetic structure factor can deviate significantly from the nuclear (positional) structure factor for magnetically interacting nanoparticle ensembles; see also Refs.~\cite{hayter82,pynn83}, where the structure factor of a magnetically saturated ferrofluid has been derived. Since correlations between the particle magnetizations are magnetic-field-dependent and also anisotropic~\cite{gazeau02,honecker2020}, extending the present Stoner-Wohlfarth approach to higher concentrations does not change the main statement of the present work, namely that the \textit{magnetic} SANS cross section of a randomly-oriented nanoparticle ensemble is, in the $\mathbf{B} \perp \mathbf{k}_0$ geometry, generally anisotropic.

\section{Conclusions}
\label{summary}

We have analyzed the angular anisotropy of the magnetic SANS cross section of spherical Stoner-Wohlfarth particles using the Landau-Lifshitz equation. Depending on the symmetry of the magnetic anisotropy of the particles (uniaxial, cubic), an anisotropic randomly-averaged magnetic SANS pattern may result in the perpendicular scattering geometry, even in the remanent or fully demagnetized state. The magnetic scattering in the parallel geometry is, as expected, isotropic. Inhomogeneously magnetized nanoparticles also generally exhibit an anisotropic randomly-averaged magnetic SANS response. Subtraction of the total scattering at saturation from data at lower fields might help to access the intrinsic anisotropy of the particles.

\appendix

\section{Magnetization curves and crosscorrelation values of spherical Stoner-Wohlfarth particles with uniaxial and cubic anisotropy}
\label{appendixA}

\begin{figure}[tb!]
\centering
\resizebox{1.0\columnwidth}{!}{\includegraphics{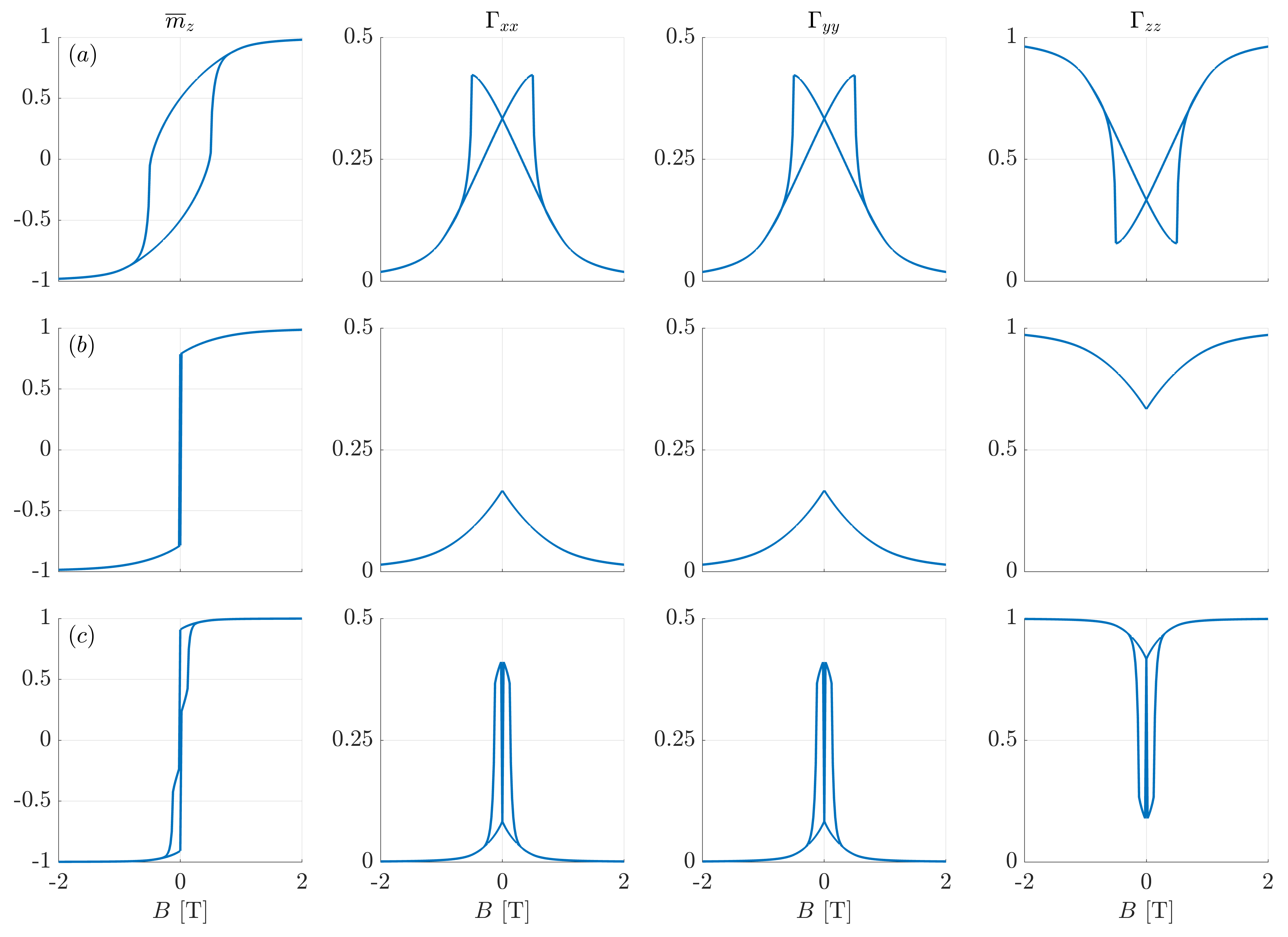}}
\caption{Hysteresis loops $\overline{m}_z(B)$ (the reduced remanences and coercivities are indicated) and second moments $\Gamma_{\alpha \alpha}$ of the components of the magnetization vectors from the Stoner-Wohlfarth model with uniaxial anisotropy. (\textit{a})~$k_{\mathrm{u}1} = +1$ and $k_{\mathrm{u}2} = 0$; (\textit{b})~$k_{\mathrm{u}1} = -1$ and $k_{\mathrm{u}2} = 0$; (\textit{c})~$k_{\mathrm{u}1} = -0.5$ and $k_{\mathrm{u}2} = +0.5$.}
\label{fig1app}
\end{figure}

\begin{figure}[tb!]
\centering
\resizebox{1.0\columnwidth}{!}{\includegraphics{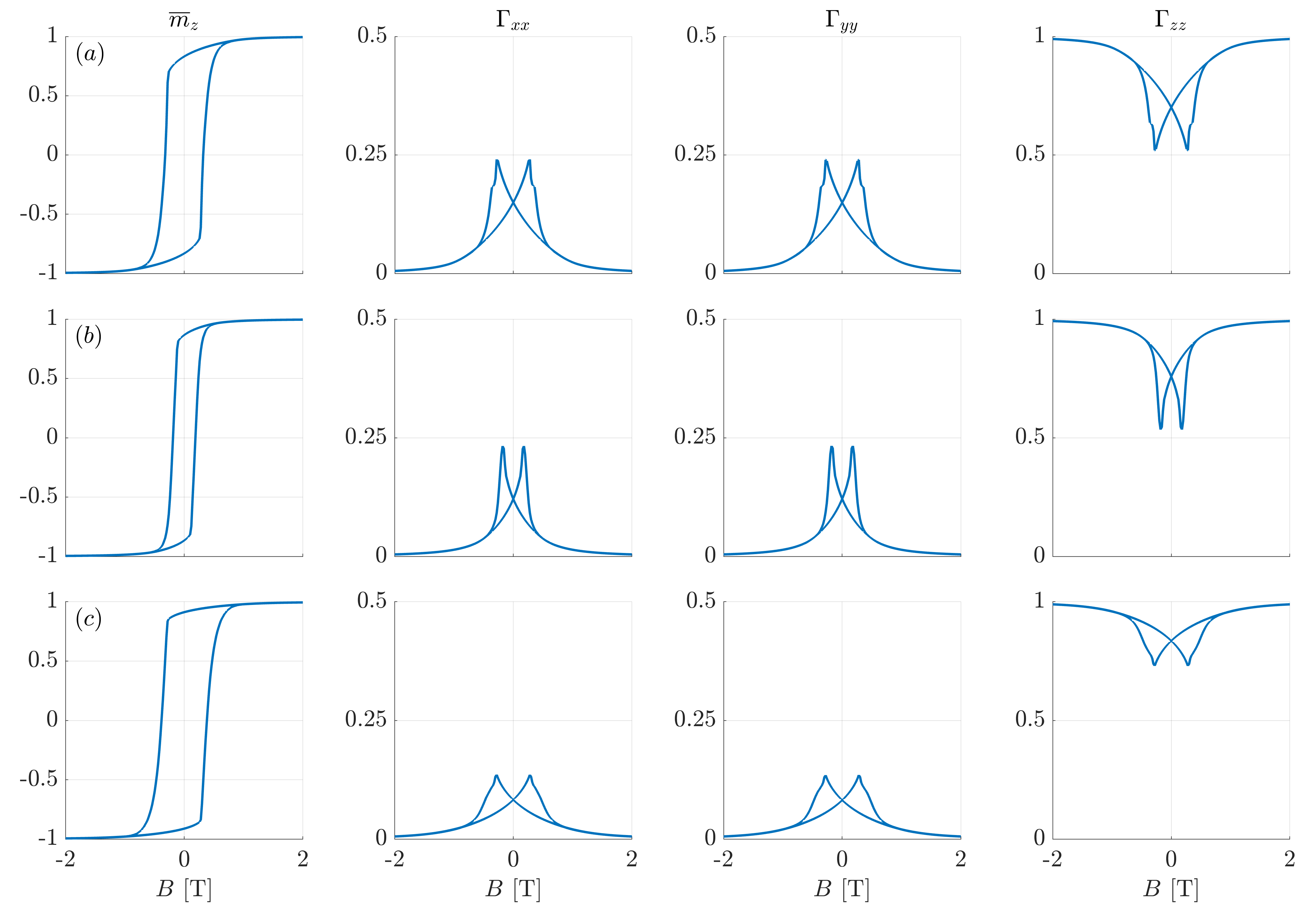}}
\caption{Similar to Fig.~\ref{fig1app}, but for cubic anisotropy. (\textit{a})~$k_{\mathrm{c}1} = +1$ and $k_{\mathrm{c}2} = 0$; (\textit{b})~$k_{\mathrm{c}1} = -1$ and $k_{\mathrm{c}2} = 0$; (\textit{c})~$k_{\mathrm{c}1} = -1$ and $k_{\mathrm{c}2} = +9$.}
\label{fig2app}
\end{figure}

Figs.~\ref{fig1app} and \ref{fig2app} show the hysteresis loops $\overline{m}_z(B)$ and the field dependencies of the second moments of the components of the magnetization vectors, $\Gamma_{\alpha \alpha}(B)$, from the Stoner-Wohlfarth model for uniaxial and cubic anisotropy. The six cases specified in Table~\ref{table1} are considered. All crosscorrelations $\Gamma_{\alpha \beta}$ with $\alpha \neq \beta$ vanish, and $\overline{m}_x(B) = \overline{m}_y(B) = 0$.

\section*{Acknowledgements}
Mathias Bersweiler and Ivan Titov are acknowledged for critically reading the manuscript.

\section*{Funding information}
We thank the National Research Fund of Luxembourg for financial support (AFR Grant No.~15639149 and PRIDE MASSENA Grant).

%


\end{document}